\newcommand*{\ATLASLATEXPATH}{}
\pdfoutput=1
\documentclass[cernpreprint,texlive=2016,txfonts,USenglish,texmf]{\ATLASLATEXPATH atlasdoc}
\usepackage{\ATLASLATEXPATH atlaspackage}
\usepackage{\ATLASLATEXPATH atlasbiblatex}
\usepackage{\ATLASLATEXPATH atlascontribute}
\usepackage{\ATLASLATEXPATH atlasphysics}
\usepackage{\ATLASLATEXPATH atlas_tables-defs}
\usepackage{siunitx}
\usepackage{subfig}
\usepackage{tauJets-defs}
\usepackage[inline]{enumitem}
\newcommand{\Ndata}{N_{\text{data}}}
\newcommand{\Nbkg}{N_{\text{bkg}}}
\newcommand{\Nbkgntt}{N_{\text{bkg-non\ttbar}}}

\newcommand{\effb}{\epsilon_{\text{FF-\ttbar}}}
\newcommand{\FF}{\ensuremath{\text{FF}}\xspace}
\AtlasTitle{Measurement of the $t\bar{t}$ production cross section in the
$\tau+{\text{jets}}$ final state in $pp$ collisions at $\sqrt{s}=8$~TeV using the ATLAS detector}
\author{ATLAS Collaboration}
\AtlasRefCode{TOPQ-2015-18}
\PreprintIdNumber{CERN-EP-2016-288}
\AtlasJournalRef{Phys. Rev. D 95, 072003 (2017)}
\AtlasDOI{10.1103/PhysRevD.95.072003}
\AtlasAbstract{%
A measurement of the inclusive $pp\to t\bar{t}+X$ production
cross section in the $\tau+{\text{jets}}$ final state
using only the hadronic decays of the \tau lepton is presented. 
The measurement is
performed using \SI{20.2}{\per \fb}  of proton-proton collision data
recorded at a center-of-mass energy of $\sqrt{s}= \SI{8}{\tev}$ with
the ATLAS detector at the Large Hadron Collider.
The cross section is measured via a counting experiment by imposing a
set of selection criteria on the identification and kinematic variables
of the reconstructed particles and jets, and on event kinematic variables
and characteristics.
The production cross section is measured to be
$\sigma_{t\bar{t}}= 239 \pm 29 $~pb, which is in agreement with the measurements in
other final states and the theoretical predictions at this
center-of-mass energy.
}
\hypersetup{pdftitle={ATLAS document},pdfauthor={The ATLAS Collaboration}}

\begin{document}

\maketitle
\tableofcontents
%
\section{Introduction}
\label{sec:intro}
%
An important component of the Large Hadron Collider (LHC)~\cite{LHC}
physics program is the measurement of the properties of the
top quark, which is the most massive fundamental particle observed
to date. With approximately one top-quark pair produced every second,
the data sample used in this analysis is significantly larger than
previously available samples
allowing for precise measurements of top-quark properties using
final states that were previously limited by their statistical
uncertainty. This
article reports on a measurement of the \ttbar production
cross section in the $\tau+\jets$ final state, where the hadronic
final states of the \tau lepton (\tauh) are used exclusively.
This measurement,
which is of comparable precision to the $\mu+\jets$ and $e+\jets$
cross section measurements by the ATLAS
Collaboration~\cite{TOPQ-2013-06}, provides a cross-check of the \ttbar
production cross section measurements in the other final states. In
addition, differences between measurements or between measurement and 
theory could lead to the
discovery of non-Standard-Model physics or to limits on its possible
extensions. Previous measurements in this final state
have been performed by the D0~\cite{Abazov:2010pa}
and CDF~\cite{Collaboration:2012hz} collaborations
at the Tevatron operating at $\sqrt{s}=\SI{1.96}{\tev}$, and by
the ATLAS~\cite{TOPQ-2012-04} and CMS~\cite{CMS-TOP-11-004} collaborations
at the LHC operating at $\sqrt{s}=\SI{7}{\tev}$.
Besides the measurement in the $\ell+\jets$ ($\ell=e,\ \mu,\ \tau$)
final state at $\sqrt{s}=\SI{8}{\tev}$,
the $\ttbar$ production cross section has also been measured
in the dilepton ($e^+e^-$, $\mu^+\mu^-$,
and $e^\pm\mu^\mp$) final state by the ATLAS and CMS
collaborations~\cite{TOPQ-2013-04, CMS-TOP-12-007}. Since the different
channels in which this measurement has been performed have different
backgrounds and systematic uncertainties, each measurement serves as
a cross-check of the others.

The final state of the process used in this measurement,
$\ttbar\to \tau+\jets$, includes one top quark decaying as
$t\to Wb\to\tau\nu_{\tau}b$ while the other decays as $t\to Wb\to qq'b$,
leading to the final-state topology of one \tau lepton,
an imbalance of momentum in the plane transverse to the beam
axis (\met), and four quark jets with two of these being $b$-quark jets.

The decay $t\to \tau\nu_{\tau}b$ provides a unique system in which
to investigate the couplings of the third-generation fermions ---
the top and bottom quarks, the \tau lepton,
and the \tau neutrino $\nu_{\tau}$ --- in a single process.
In the framework of the
Standard Model (SM), the branching ratio (BR) of the top quark decaying
to a $W$
boson and a $b$ quark is approximately $100\%$. Hence, the final state
is determined by the SM BRs of the $W$ boson, which are well
measured~\cite{Agashe:2014kda}. 
In the SM, electroweak symmetry-breaking introduces mass- and
flavor-dependent couplings. Since the top quark is the most massive quark
and the \tau lepton the most massive lepton, these fermions along with
the $b$ quark have the largest Yukawa couplings to the Higgs boson
and, hence, could lead to non-SM mass- or flavor-dependent
couplings that can change the top-quark decay rate into
final states with $\tau$ leptons. 
Therefore, any observed deviation in the
BR of $t\to\tau\nu_{\tau}b$ from that predicted by the SM would
be an indication of non-SM physics. For example, in
Type-2 two-Higgs-doublet models (2HDM)~\cite{THDM}, such as required
by the Minimal Supersymmetric Standard Model~\cite{MSSM}, the
top quark can have a significant BR to a charged Higgs boson ($H^\pm$) 
and a $b$ quark if $m_{H^\pm}<\mtop-m_b$.
For large values of $\tan\beta$, the ratio of the vacuum expectation
values of the two Higgs doublets, the charged Higgs boson preferentially
decays to $\tau\nu_{\tau}$. This thereby increases the
BR of $t\to\tau\nu_{\tau}b$ relative to the SM prediction and
leads to a larger measured value of 
$\sigma_{\ttbar}\times\text{BR}(\ttbar
\rightarrow \tau+\text{jets})$~\cite{ttbtheory1, ttbtheory2, ttbtheory3}.
Small values
of $\tan\beta$, however, would decrease the number of
$\ttbar\to\tau+\text{jets}$ events relative to the SM prediction.

The 2HDM can also produce an excess of $t\to\tau+X$ decays
if flavor-changing neutral couplings are allowed as in
Type-3 models~\cite{Cheng:1987rs, Hou:1991un}.
For example, this allows $t\to cH$ and if the Higgs boson decays as
$H\to\tau^+\tau^-$,
an excess of events with $t\to\tau+X$ decays would be observed relative to the
SM. The SM predicts $\BR(t\to cH)\approx 10^{-15}$~\cite{AguilarSaavedra:2004wm},
whereas Type-3 models predict $\BR(t\to cH)$ to be as
large as $10^{-3}$~\cite{Chen:2013qta,Atwood:2013ica,AguilarSaavedra:2004wm}. 

This article presents an analysis using the  $\tauh+\jets$ final state to
measure the \ttbar production
cross section in $\sqrt{s} = \SI{8}{\tev}$ proton-proton ($pp$) collisions.
The data sample for this measurement
was recorded using the ATLAS detector and corresponds to
an integrated luminosity of \SI{20.2}{\per\fb}.
The ATLAS detector is briefly described in Sec.~\ref{sec:detector}.
Section~\ref{sec:samples} presents the data and
simulated event samples used in this measurement. 
The reconstruction of jets, \tau leptons, and missing transverse momentum is
discussed in Sec.~\ref{sec:objReco}. The event selection is described in
Sec.~\ref{sec:eventS} and the methods used to estimate the backgrounds
are discussed in Sec.~\ref{sec:mJet}. The calculation of the
production cross section is given in Sec.~\ref{sec:Xsect} and
the estimation of the various systematic uncertainties is
presented in Sec.~\ref{sec:systUn}. The results 
of the analysis and the interpretations are discussed in
Sec.~\ref{sec:result}. Finally, the analysis is summarized in
Sec.~\ref{sec:conclusion}. 
\section{ATLAS Detector}
\label{sec:detector}
%
\newcommand{\AtlasCoordFootnote}{%
ATLAS uses a right-handed coordinate system with its origin at the nominal interaction point (IP)
in the center of the detector and the $z$-axis along the beam pipe.
The $x$-axis points from the IP to the center of the LHC ring,
and the $y$-axis points upwards.
Cylindrical coordinates $(r,\phi)$ are used in the transverse plane, 
$\phi$ being the azimuthal angle around the $z$-axis.
The pseudorapidity is defined in terms of the polar angle $\theta$ as $\eta = -\ln \tan(\theta/2)$.
Angular distance is measured in units of $\Delta R \equiv \sqrt{(\Delta\eta)^{2} + (\Delta\phi)^{2}}$.}

\label{sec:atlas1}
The ATLAS detector~\cite{PERF-2007-01} at the LHC covers nearly the entire solid angle around the collision point.
It consists of an inner tracking detector surrounded by a thin superconducting solenoid, electromagnetic and hadronic calorimeters,
and a muon spectrometer incorporating three large superconducting toroid magnets.
The inner detector (ID) is immersed in a \SI{2}{\tesla} axial magnetic field 
and provides charged-particle tracking in the range $|\eta| < 2.5$, where $\eta$ is
the pseudorapidity of the particle.\footnote{\AtlasCoordFootnote}

The high-granularity silicon pixel detector covers the interaction region and typically provides three position measurements per track.
It is followed by the silicon microstrip tracker, which usually provides four two-dimensional measurement points per track.
These silicon detectors are complemented by the transition radiation tracker,
which enables radially extended track reconstruction up to $|\eta| = 2.0$. 
The transition radiation tracker also provides electron identification information 
based on the fraction of hits 
above a higher energy-deposition threshold corresponding to transition radiation.

The calorimeter system covers the pseudorapidity range $|\eta| < 4.9$.
Within the region $|\eta|< 3.2$, electromagnetic calorimetry is provided by barrel and 
endcap high-granularity lead/liquid-argon (LAr) electromagnetic calorimeters,
with an additional thin LAr presampler covering $|\eta| < 1.8$
to correct for energy loss in material upstream of the calorimeters.
Hadronic calorimetry is provided by the steel/scintillator-tile calorimeter,
segmented into three barrel structures within $|\eta| < 1.7$, and two copper/LAr hadronic endcap calorimeters.
The solid angle coverage is completed with forward copper/LAr and tungsten/LAr calorimeter modules
optimized for electromagnetic and hadronic measurements, respectively.

The muon spectrometer comprises separate trigger and
high-precision tracking chambers measuring the deflection of muons in a magnetic field generated by superconducting air-core toroids.
The precision chamber system covers the region $|\eta| < 2.7$ with three layers of monitored drift tubes,
complemented by cathode strip chambers in the innermost layer of the forward region, where the background is highest.
The muon trigger system covers the range $|\eta| < 2.4$ with resistive plate chambers in the barrel, and thin gap chambers in the endcap regions.

A three-level trigger system is used to select interesting events~\cite{PERF-2011-02}.
The Level-1 trigger is implemented in hardware and uses a subset of detector information
to reduce the event rate to a design value of at most \SI{75}{\kHz}.
This is followed by two software-based trigger levels that together reduce the event
rate to about \SI{400}{\Hz}.
\section{Data and Simulation Samples}
\label{sec:samples}
The $pp$ collision data sample used in this measurement was collected with
the ATLAS detector at the LHC
and corresponds to the full \SI{20.2}{\ifb} of
integrated luminosity collected at this energy with the
requirement of  
stable beam conditions and an operational detector.

In order to estimate the effects of detector resolution and acceptance
on signal and background, and to estimate the backgrounds, a full
\textsc{Geant}4-based detector simulation is
utilized~\cite{Agostinelli:2002hh, SOFT-2010-01}. In addition,
to estimate the modeling uncertainties of the various physics processes
in an efficient manner, a detector simulation using 
parameterized calorimeter showers is also used~\cite{ATL-PHYS-PUB-2010-013}.
To account for an average of 20.7 interactions per bunch crossing,
$pp$ interactions are generated using
\PYTHIA~v8.165~\cite{Sjostrand:2006za, Sjostrand:2007gs} and overlaid on
the signal and
background Monte Carlo (MC) simulation samples in accordance with the
average observed number of interactions per bunch crossing. 
All simulated samples are reconstructed and analyzed with the same
algorithms and techniques as for the recorded $pp$ collision data. Only
events with at least one charged lepton ($e$, $\mu$, \tau) in the final
state are generated. 

To estimate the acceptance of the event selection for \ttbar events,
several MC samples
are generated with the top-quark mass set to $\mtop=\SI{172.5}{\gev}$.
The nominal sample is generated using the next-to-leading-order (NLO)
matrix element (ME) event generator
\POWHEG-\textsc{box}~\cite{Frixione:2007nw,Nason:2004rx, powheg, Alioli:2010xd}
with the CT10~\cite{pdfCT10} NLO parton distribution functions (PDF).
The output of \POWHEG-\textsc{box} is then processed by
\PYTHIA~v6.426~\cite{Sjostrand:2006za} to perform the parton showering (PS),
hadronization, and generation of
the underlying event (UE). For the UE generation to agree with data,
\PYTHIA~v6.426 uses the leading-order (LO) CTEQ6L1 PDF
set~\cite{pdfCTEQ} and a set of tuned parameters referred to as
the Perugia 2011C tune~\cite{perugiatune}.
To regulate high-\pt radiation in \POWHEG-\textsc{box} and provide
ME/PS matching, the resummation
damping factor \hdamp is set to $\mtop$~\cite{TOPQ-2012-03}.
The \ttbar sample is normalized using the theoretical production
cross section, which for $pp$ collisions at $\sqrt{s} = \SI{8}{\tev}$
is $\sigma_{\ttbar}= 253^{+13}_{-15}$~\si{pb}
assuming a top-quark mass of \SI{172.5}{\gev}. It has been calculated at
next-to-next-to leading order (NNLO) in \alphas including resummation of
next-to-next-to-leading logarithmic (NNLL) soft gluon terms with
top++2.0~\cite{Cacciari:2011hy, Beneke:2011mq, Baernreuther:2012ws,
Czakon:2012zr, Czakon:2012pz, Czakon:2013goa, Czakon:2011xx}.
The systematic uncertainty in the cross section due to the uncertainties
in the PDF and $\alphas$ is calculated using the PDF4LHC 
prescription~\cite{Botje:2011sn} with the MSTW2008 68\% CL
NNLO~\cite{Martin:2009iq, Martin:2009bu}, CT10
NNLO~\cite{pdfCT10, Gao:2013xoa}, and NNPDF2.3 five~flavor
number~\cite{Ball:2012cx} PDF sets and added in quadrature to the
uncertainties due to the renormalization and factorization scales. 

Systematic uncertainties associated with the \ttbar modeling
are evaluated using alternative sets of simulated events that are
compared to the nominal sample, with the nominal and alternative sets
processed using the parameterized detector
simulation~\cite{ATL-PHYS-PUB-2010-013}. 
Since the choice of the ME event generator can affect the estimate of the acceptance,
the ME event generator \MCatNLO~v4.01~\cite{mcnlo} and the PS/UE simulator
\HERWIG~v6.520~\cite{herwig}, with \JIMMY~v4.31~\cite{Butterworth:1996zw} is compared to the
\POWHEG-\textsc{box}~\cite{powheg} event generator where the PS is simulated by
\HERWIGNS+\JIMMY. The effect of the PS and hadronization models on the acceptance is
investigated by comparing the \POWHEGNS+\PYTHIA event generator with
$\hdamp=\infty$ to the \POWHEGNS+\HERWIG event generator.
Finally, the effect of initial- and
final-state radiation (ISR and FSR) is estimated using two \ttbar samples
generated in the same manner as the nominal sample, but with the renormalization and factorization
scales multiplied by 2.0 (0.5), the regularization parameter \hdamp set to
$\mtop$ ($2\mtop$), and using the Perugia 2012 radLo (radHi) UE tune,
giving less (more) radiation. Table~\ref{tab:mc_gens_signal}
summarizes the samples used to calculate the systematic
uncertainties for the \ttbar process.   

A variety of MC event generators are used to simulate the backgrounds
containing charged leptons in the final state, which are summarized in
Table~\ref{tab:mc_gens_bkg}. 
Vector-boson production with additional jets
($pp\to V+\jets$, with $V=W,\ Z$ and two to seven jets) is simulated
using the LO parton-level ME event generator \ALPGEN~\cite{alpgen}
with the PS/UE
generated by \PYTHIA~v6.426, as for the nominal \ttbar samples. In order to
avoid double counting, final states generated by the LO parton-level
event generator \ALPGEN and the parton-level shower evolution of \PYTHIA,
the MLM matching algorithm is used~\cite{Hoche:2006ph}. The matching
algorithm is applied inclusively to the $V+5$ light-parton events
and exclusively to the other events. Associated production of vector bosons
with heavy-flavor partons ($V+c+\jets$, $V+\ccbar+\jets$, $V+\bbbar+\jets$)
is simulated separately. Inclusive $V+\jets$
samples are formed by combining the light- and heavy-quark samples
according to their respective cross section. An overlap removal scheme
is used to avoid double counting the contribution of additional
heavy flavor partons.
The cross sections used to normalize the samples are calculated at
NNLO~\cite{Hamberg:1990np, Anastasiou:2003ds}.

Electroweak production of the top quark (single-top) is simulated using
\POWHEG-\textsc{box}~\cite{Alioli:2009je} and \PYTHIA~v6.426 with the
CT10 PDF set.
The MC sample for the $t$-channel process is normalized
using the NNLO calculation in Ref.~\cite{Kidonakis:2011wy} while
the $s$-channel sample is normalized with the
$\text{NNLO}+ \text{NNLL}$ cross section in Ref.~\cite{Kidonakis:2010tc}
and the $Wt$ channel is normalized with the 
$\text{NNLO}+ \text{NNLL}$ calculation in
Ref.~\cite{Kidonakis:2010ux}. In order to remove the overlap with
\ttbar production, the $Wt$ sample is produced using the ``diagram
removal'' generation scheme~\cite{Frixione:2008yi}.

In addition, diboson ($WW$, $WZ$) production samples are generated
using \HERWIG with the CTEQ6L1 PDF set. These samples are normalized using the
NLO calculation in Ref.~\cite{Campbell:2011bn}.

\begin{table}[htbp]
  \begin{center}
    \renewcommand{\arraystretch}{1.1}
    \begin{tabular}{lccc}
    \hline
     Systematic uncertainty & Generator & Parton shower & Tune set  \\
      \hline
     Nominal            & \POWHEG & \PYTHIA & Perugia2011C  \\
     Parton shower      & \POWHEG & \HERWIG & AUET2 \\
     Generator          & \MCatNLO & \HERWIG & AUET2 \\
     ISR/FSR            & \POWHEG & \PYTHIA & Perugia2012radLo \\ 
     ISR/FSR            & \POWHEG & \PYTHIA & Perugia2012radHi \\ 
     \hline
    \end{tabular}
    \caption{List of the \ttbar MC samples used in studying the modeling
      uncertainties. The PDF set used for all event generators
      is CT10.
    }
    \label{tab:mc_gens_signal}
  \end{center}
\end{table}

\begin{table}[htbp]
  \begin{center}
    \renewcommand{\arraystretch}{1.1}
    \begin{tabular}{lcccc}
      \hline
      Process & Generator & Parton shower & PDF set & Tune set\\
      \hline
      $W$+\jets                 & \ALPGEN & \PYTHIA & CTEQ6L1 &
      Perugia2011C \\  
      $Z$+\jets                 & \ALPGEN & \PYTHIA & CTEQ6L1 & Perugia2011C \\  
      Single top ($Wt$-channel)& \POWHEG & \PYTHIA & CT10 & Perugia2011C \\
      Dibosons ($WW$,$WZ$,$ZZ$)& \HERWIG & \HERWIG & CTEQ6L1
      & AUET2B \\
      \hline
    \end{tabular}
    \caption{The matrix element event generators and the parton shower
      simulators used to generate the MC simulated background events. 
      The parton distribution functions used by the event
      generators and
      the set of tuned parameters used in the parton shower simulators
      are also shown.}
    \label{tab:mc_gens_bkg}
  \end{center}
\end{table}
\section{Object Reconstruction}
\label{sec:objReco}
The final state in this measurement contains four quark jets of which two are
$b$-quark jets, a $W$ boson decaying to a neutrino and a \tau lepton that decays
to hadrons (\tauh) and a neutrino.
Jets are reconstructed using the anti-\kt
algorithm~\cite{Cacciari:2005hq,Cacciari:2008gp} with the radius
parameter set to $R=0.4$. To account for inhomogeneities and the
noncompensating response of the calorimeter, the reconstructed jet energies
are corrected through \pt- and \eta-dependent factors that are derived
in MC simulation and validated in data. Any remaining discrepancies
in the jet energy scale are calibrated using an {\em in situ} technique where
a well-defined reference object is momentum-balanced with a
jet~\cite{PERF-2012-01}. To ensure that jets originate from the vertex
that produced the event, 
the fraction of the scalar \pt sum of all tracks matched
to the jet and originating
at this vertex (jet vertex fraction) to the scalar \pt sum of all
tracks associated with this jet but originating from any vertex must
be $> 0.5$ for jets with $\et< \SI{50}{\gev}$ and $|\eta|<2.4$.

To identify jets initiated by $b$ quarks ($b$-tagging), a multivariate
algorithm
is employed~\cite{PERF-2012-04}. This algorithm uses the impact parameter
and reconstructed secondary vertex information of the tracks contained in
the jet as input for a neural network. Jets initiated by $b$ quarks are
selected
by setting the algorithm output threshold such that a 70\%
selection efficiency is achieved in simulated \ttbar events
with a 1\% misidentification rate for light-flavor jets.
Since the $b$-quark
selection efficiency differs between data and MC simulation, \pt dependent
correction
factors are derived to correct for this difference~\cite{PERF-2012-04}. 
These correction factors differ from unity by less than 3\% over the 
entire \pt range.

Decays of the \tau~lepton into hadrons and a neutrino are classified as
either single prong (\tauone), where the \tau~lepton decays to a single charged
particle, or three prong (\tauthree),
where the decay products are three charged particles with a net unit charge,
and for each classification zero or more \pizero mesons can be present.
Identification of a \tauh begins with a reconstructed jet, as described
above, having $\pt>\SI{10}{\gev}$ and $|\eta|<2.5$.
The \tauh classification is achieved by counting the number of tracks
with $\pt>\SI{1}{\gev}$ in a cone of size $\Delta R=0.2$ around the jet axis.
To discriminate against quark- or gluon-initiated jets, a set of discriminating
variables is used to train a multivariate Boosted Decision Tree (BDT)
separately for single-prong and three-prong \tau decays
using \tauh from simulated samples of vector-bosons decaying
into \tau leptons that cover the kinematic range expected in data
and a background sample
enriched in dijet events from data~\cite{PERF-2013-06}. Three categories
of discriminating variables are used. The first category comprises
those variables that apply to all candidates.
These are associated with the jet shape in both the tracking
system and
calorimeter. The second category are those variables that apply
only to the
single-prong \tau lepton decays. These include
the impact parameter significance and the number of tracks in an
isolation region ($0.2<\Delta R<0.4$) around the jet axis. The third
and final category are those that apply to the three-prong
\tau lepton decays. These variables include the
decay length significance in the transverse plane, the invariant mass
of the reconstructed tracks, and the maximum track separation ($\Delta R$)
from the jet axis. An additional set of 
variables is used for those \tauh containing \pizero mesons. These include
the number of \pizero mesons, the invariant mass of the tracks plus \pizero
mesons, and the ratio of track plus \pizero \pt to the calorimeter
energy only measurement.
Furthermore, any jet that satisfies $\Delta R < 0.2$
of a \tauh is removed. In addition, a BDT that
includes discriminating variables against
electrons is trained to reduce the electron contamination for
the \tauone candidates.
Low \pt muons that stop in the calorimeter and overlap with energy deposits
from other sources can mimic a \tauh. These are characterized by a large
fraction of energy deposited in the electromagntic calorimeter and
a small ratio of track-\pt to calorimeter-\ET. Muons that produce large
energy deposits in the calorimeter can also be misidentified as a \tauh.
These are characterized by a small fraction
of energy in the electromagnetic calorimeter and a large track-\pt to
calorimeter-\ET ratio.
Strict selection requirements based on the two variables described
are applied to avoid muons being misidentified as \tauh.
In addition, the reconstructed four-vector of the \tauh candidate is
not corrected for the unobserved neutrino kinematics.

Since undetected neutrinos occur in the final state, a momentum
imbalance in the transverse plane is expected. The missing transverse
momentum (\met) is calculated as the negative of the vector sum of
the transverse momentum of all reconstructed objects
and of the calorimeter energy deposits not associated to any
reconstructed object after the appropriate energy
corrections have been applied~\cite{PERF-2011-07}.
\section{Event Selection}
\label{sec:eventS}
Events are selected that satisfy the $\met>\SI{80}{\gev}$
trigger with an offline reconstruction
requirement of $\met> \SI{150}{\gev}$. This is the point
at which the trigger has almost reached full efficiency.
Furthermore, events are required to contain
a hard collision primary vertex with at least four associated charged
particle tracks of $\pt>\SI{0.4}{\gev}$.
If there are multiple primary vertices in an event, the one with the
largest sum of track $\pt^2$ is selected.
To reduce contamination from events with $\ttbar\to e(\mu)+\jets$,
an event is rejected if it contains an electron
($\pt^{e}>\SI{25}{\gev}$)~\cite{PERF-2013-03}
or a muon ($\pt^{\mu}>\SI{20}{\gev}$)
candidate~\cite{PERF-2014-05}, each with $|\eta|<2.5$ that satisfy
the corresponding selection in the ATLAS $\ttbar\to e(\mu)+\jets$
cross-section measurement~\cite{TOPQ-2013-06}.
The event must also contain at least two jets with $\et>\SI{25}{\gev}$
and $|\eta|< 2.5$. In addition,
at least two of the jets in the event must be identified as $b$-quark jets
using a $b$-tagging requirement with 70\% efficiency.
Each event is also required to contain
at least one \tauh that decays to either one or three charged particles
with $\et> \SI{20}{\gev}$ and $|\eta|< 2.5$ and a \tau lepton identification
requirement that discriminates against quark and gluon initiated jets
such that
the efficiency is 40\% for single-prong and 35\% for three-prong \tauh
with a rejection factor between 100 and
1000 depending on the \pt and \eta for each. 
This identification requirement defines the standard \tauh selection.
Since the background for \tauone and
\tauthree identification is different, the two samples
are analyzed separately with the \tauone (\tauthree) analysis requiring
one or more \tauone (\tauthree) only. In each case, the highest-\pt
\tauh is used with less than $1\%$ of events containing more than a single
$\tauh$. The combined result is produced by requiring an event
to contain either one or more \tauone or \tauthree and selecting the
highest-\pt \tauh in the event.
In order to preferentially
select events where the \tauh and \met originate from $W$-boson decays,
the transverse mass
is required to satisfy $\mT<\SI{90}{\gev}$, where \mT comprises
the \tauh with the largest \pt and the value of the \met of the event. The
square of the transverse mass is
defined as $\mT^2=\pt{^{\tauh}}\met[1-\cos\Delta\phi(\tauh,\met)]$
and $\Delta\phi(\tauh,\met)$ is the azimuthal angle between the
direction of the \tauh and the \met of the event.
\section{Background Estimation}
\label{sec:mJet}
To determine the number of $pp\to\ttbar+X\to\tauh+\jets$ events
in the data sample, estimates of the various backgrounds are subtracted. These
originate from two sources: the
backgrounds with real \tauh and those with misidentified \tauh
in the final state.
The backgrounds containing real \tauh in the final state include
single-top-quark events, $V+\jets$ events,
and diboson events. These backgrounds are estimated in simulation and
normalized using their theoretical cross sections as discussed in
Sec.~\ref{sec:samples}.
The misidentified (fake) \tauh background consists of  events
from processes where a charged lepton
($e^\pm$ and $\mu^\pm$) is misidentified as a \tauh and multijet events
that have a mismeasured \met
and a quark- or gluon-initiated jet that is misidentified as a \tauh.

Misidentification of electrons and muons as \tauh is significantly
reduced by
applying the selection criteria discussed in Sec.~\ref{sec:objReco}.
The \ttbar background where an electron or a muon is
misidentified as a \tauh is simulated in \POWHEG+\PYTHIA and
normalized using the theoretical \ttbar production cross section.
The contribution from other
processes where an electron or a muon is misidentified
as a \tauh is found to be negligible.

To estimate the fraction of events in
which a jet is misidentified as a \tauh, a data-based method is
used where this fraction is evaluated in a control sample that is
divided into two components: one with the standard \tauh selection
and the other with an inverted \tauh selection.
The transfer factor is the ratio of the number of events with
misidentified \tauh in the nominal sample to that in the inverted sample.
This transfer factor, which is referred as the fake-factor \FF,
is then applied to the signal sample with the inverted \tauh selection,
which yields
the fraction of misidentified \tauh in the signal sample with the
nominal \tauh selection.
The inverted \tauh selection is determined such that
the fraction of quark-
and gluon-jets that can be misidentified as a \tauh is similar to the
fractions when the standard \tauh selection is applied,
as derived from MC simulation.
All other requirements are the same as for the signal sample.
This technique, known as the fake-factor method, has
been used in previous ATLAS measurements~\cite{EXOT-2014-05}.

To ensure a large fraction of events with jets misidentified as \tauh,
the control sample is required to satisfy
a muon trigger with only a single
reconstructed muon satisfying the requirement $\pt>\SI{25}{\gev}$
and $|\eta|<2.5$.
In addition, each event is also required to satisfy
the following criteria:
\begin{enumerate*}[{label=(\arabic*)}]
  \item contain a primary vertex with at least four associated tracks;
  \item contain at least two jets and no jet in the event satisfying
    the $b$-jet criteria;
  \item contain a single \tauh satisfying selection criteria
    that are less restrictive than the nominal.
\end{enumerate*}
The control sample is then separated into a component satisfying
the standard \tauh identification and a second component that
satisfies the inverted identification criteria. This set of selections
ensures that the
control sample is enriched with misidentified \tauh for both the
standard and the inverted \tauh identification criteria. The number of
data events selected with the standard \tauh identification is
$28\,397$, where the contribution from real \tauh is 38\% as
estimated from simulation. For the inverted \tauh identification, the number
of data events is $84\,975$ with a contribution
of 9\%  from real \tauh. The transfer factor is calculated
in bins of
\pt and \eta after the real \tauh contributions are subtracted.
The \FF averaged over the full kinematic range of this measurement
has a value of $0.23\pm 0.01~\text{(stat.)}$.

To extract the number of misidentified \tauh in the signal sample,
the nominal selection with the inverted \tauh identification is applied
to data.
To correct for real \tauh in this sample, an estimate of the number of
real \tauh is derived from simulation and subtracted from this sample.
Next,
the derived \FF is applied to the resulting data sample according to the \pt
and \eta of the selected \tauh taking into account the number of \tauh
in the event. This yields the number of misidentified \tauh in the signal
sample.

\begin{figure}[htb!]
\subfloat[]{
\includegraphics[width=0.47\textwidth]{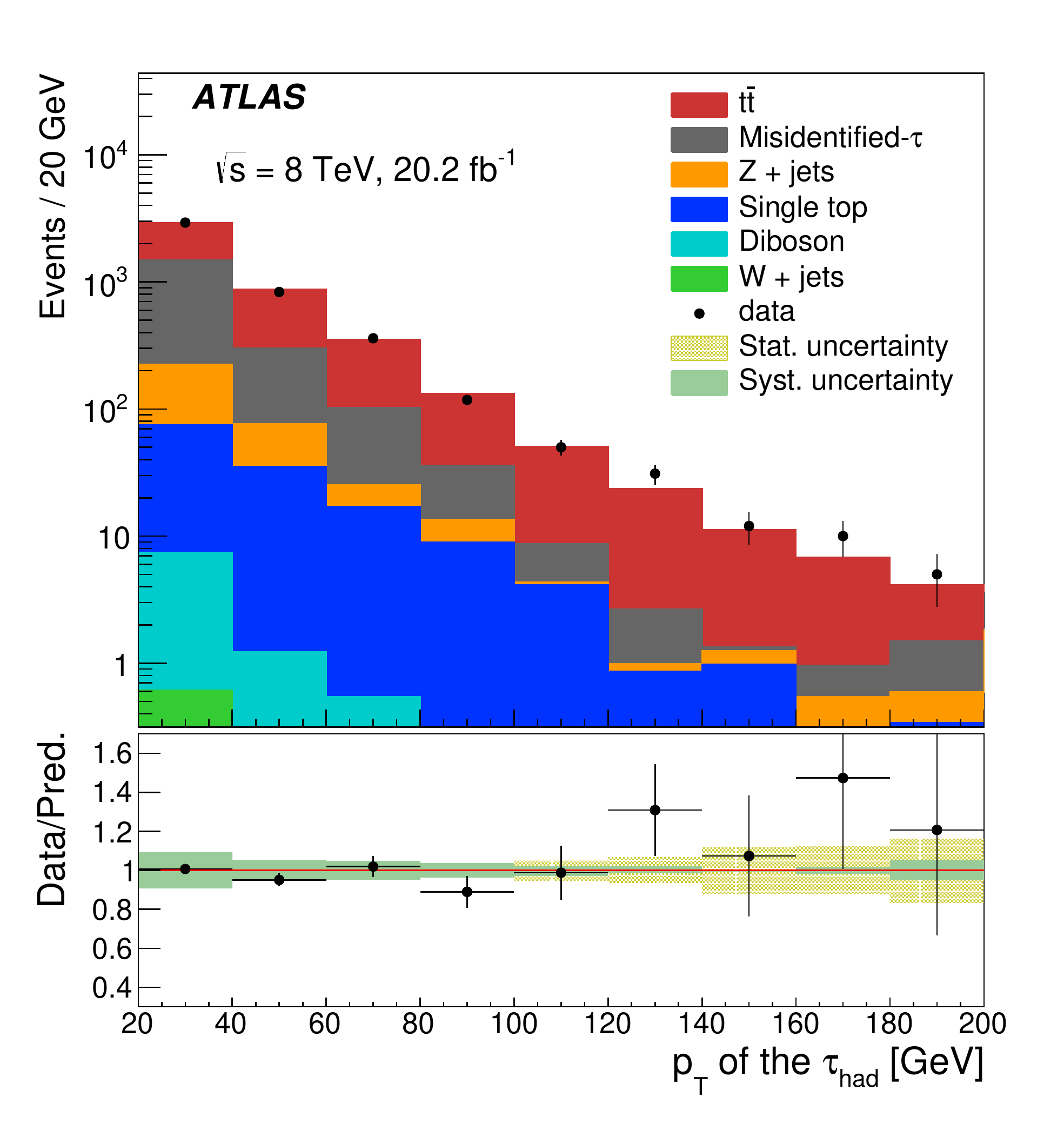}
\label{fig:fake}
}
\subfloat[]{
\includegraphics[width=0.47\textwidth]{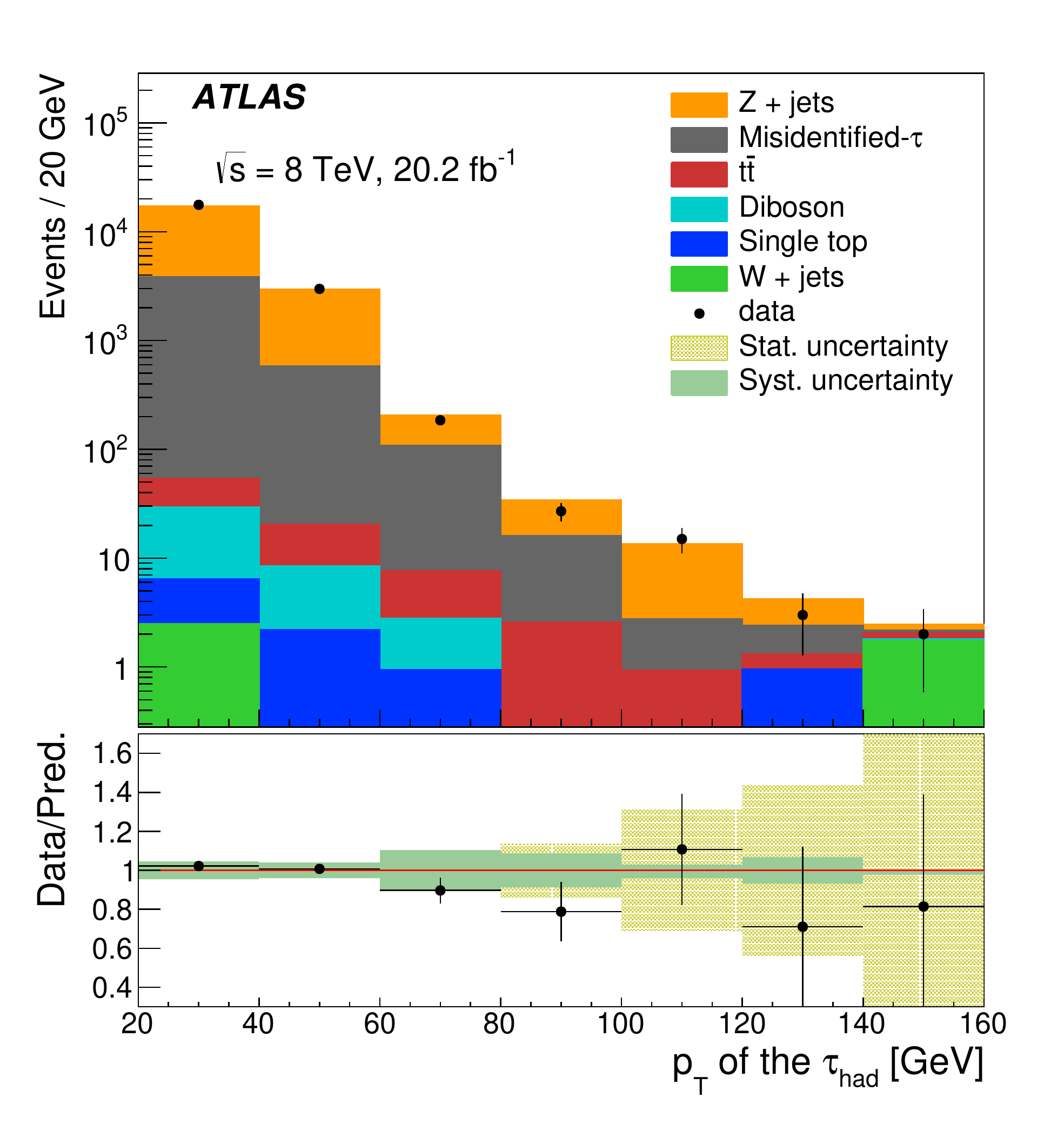}
\label{fig:real}
}
\caption{The transverse momentum distribution of the \tauh:
  \protect \subref{fig:fake}  in the $\ttbar \to \mu\tau+X$ sample dominated by
  misidentified \tauh, and \protect \subref{fig:real} in the $Z \to \tau \tau \to \mu \tau + X$
  sample dominated by real
  \tauh. The lower portion of each plot shows the ratio of the data
  over prediction, illustrating
  the level of agreement achieved between the data
  and the predicted backgrounds including the estimated number of
  misidentified \tauh.}
\label{fig:ff_validate}
\end{figure}

In order to validate this procedure, the derived \FF is applied to
a data set that does not overlap with the nominal analysis sample.
Each event
in the sample is required to satisfy a single-muon trigger and contain only
one reconstructed muon of $\pt>\SI{25}{\gev}$. In addition, a single
\tauh satisfying the
same criteria as the signal sample is required. The validation is
performed for different jet multiplicities and numbers of
$b$-quark jets by dividing the sample into the following six categories:
\begin{enumerate*}[{label=(\arabic*)}]
  \item two inclusive jets;
  \item three inclusive jets;
  \item four inclusive jets;
\end{enumerate*}
with each listed jet multiplicity containing either zero $b$-quark jets
or at least one $b$-quark jet.
Good agreement between the data and the
background estimate is seen in all categories.
An additional validation sample that is dominated by real
\tauh is
formed by selecting $Z\to\tau^+\tau^-$ events, where one \tau
lepton decays to a final state containing a $\mu$ and the
other containing hadrons. This sample is
selected by requiring:
\begin{enumerate*}[{label=(\arabic*)}]
  \item $\cos\Delta\phi(\mu,\met)+\cos\Delta\phi(\tauh,\met)>-0.15$;
  \item $\Delta\phi(\mu,\tauh)>2.4$;
  \item $m_{\text{T}}^\mu<\SI{50}{\gev}$, where $m_{\text{T}}^\mu$ is the transverse
    mass of the $\mu$ and the \met of the event;
  \item $(42<m(\mu,\tauh)<82) \si{\gev}$, the invariant mass of
    the $\mu$-$\tauh$ system;
  \item $(25<\pt^\mu<40)\si{\gev}$.
\end{enumerate*}
Figure~\ref{fig:ff_validate}  shows an example of a comparison
between the data and the prediction for regions dominated by
misidentified and real \tauh.

\section{Extraction of the \ttbar Production Cross Section}
\label{sec:Xsect}
In order to determine the \ttbar cross section,
the estimated background,
given in Table~\ref{tab:yields}, is subtracted from the
number of recorded events after the event selection is
applied, then normalized to the integrated
luminosity $\int\mathcal{L}(t) dt$ and corrected by the efficiency
$\efftt=5\times 10^{-4}$,
which is calculated from the fraction of events satisfying the geometric,
kinematic, trigger, and object identification selection, and the
effects of the detector reconstruction. Therefore, the cross section is
given as
\begin{equation}\label{eq:sigma}
  \sigma(pp\to\ttbar+X)
  =\frac{\Ndata-\Nbkg}{\BR \times\efftt\times\int\mathcal{L}(t)dt}.
\end{equation}
Furthermore, 
since the calculated efficiency corresponds to all \ttbar final
states containing leptons only, the $\BR(\ttbar\to\ell+X)=0.54$ is used.
The number of background events ($\Nbkg$) comprises
backgrounds with real \tauh that are estimated from the simulated
samples and events containing a
misidentified \tauh that is estimated using the fake-factor method
discussed in Sec.~\ref{sec:mJet}. As also discussed in 
Sec.~\ref{sec:mJet}, to estimate the  number of misidentified \tauh,
the real \tauh contribution must be subtracted including those from
\ttbar events.
Since this would require the use of the \ttbar cross section, which
is the quantity being measured, Eq.~(\ref{eq:sigma}) is reformulated as
\begin{equation}
  \label{eq:sigmaMod}
  \sigma(pp\to\ttbar+X) =
  \frac{\Ndata-\Nbkgntt}{\BR \times  (\efftt-\effb)\times\int\mathcal{L}(t)dt},
\end{equation}
where $\Nbkgntt$ represents the backgrounds estimated from the simulated
samples and the misidentified \tauh component estimated using the
fake-factor method but excludes the
subtraction of the \ttbar component. 
The efficiency $\effb=7\times 10^{-5}$ represents \ttbar events
satisfying the inverted \tauh identification.

\begin{table}[htbp]
  \centering  
\caption{\label{tab:yields}
    The number of events observed in data and obtained from simulation
    along with the
    associated statistical uncertainty for background and expected signal
    processes for the different \tauh types and the combined sample.
    The \tauone (\tauthree) samples require all \tauh in an event to be
    of that type, while the combined sample can have either \tauh type.
  }
  \label{tab:events}

  \renewcommand{\arraystretch}{1.4}
  \begin{tabular}{lrrr}
    \toprule
    Event counts & \tauone & \tauthree & \tauh\\
    \midrule
    $\ttbar \rightarrow e/\mu+\jets $ & $21.8 \pm 4.7$ & $6.8 \pm 2.5$ & $28.3 \pm 5.3$\\  
    Single top & $107 \pm 10$ & $33.9 \pm 5.8$ & $141 \pm 12$\\
    $W+\jets$ & $71.7 \pm 8.5$ & $27.1 \pm 5.2$ & $99 \pm 10$\\
    $Z+\jets$ & $7.2 \pm 2.7$ & $1.6 \pm 1.3$ & $8.7 \pm 3.0$\\
    Diboson & $1.0 \pm 1.0$ & $0.4\pm 0.6$ & $1.5 \pm 1.2$\\
    Misidentified-\tauh & $46.6 \pm 6.8$ & $24.9 \pm 5.0$ & $74.9 \pm 8.7$\\
    Expected $\ttbar \rightarrow \tau+\jets$ & $1084 \pm 33$
                                             &$312\pm 18$ & $1398 \pm 37$\\
    \midrule
    Total Expected &$1339\pm 37$ & $407\pm 20$ & $1751\pm 42$\\
    \midrule
    Data & \multicolumn{1}{l}{1278}  & \multicolumn{1}{l}{\ 395} 
         & \multicolumn{1}{l}{1678}\\
    \bottomrule
  \end{tabular}
\end{table}
\section{Systematic Uncertainties}
\label{sec:systUn}
Systematic uncertainties are
grouped into those pertaining to object identification along with its energy
and momentum measurement, theoretical modeling, background evaluation,
and the luminosity.
The systematic uncertainties are evaluated by performing a variation
of each parameter related to the associated quantity and propagating
the overall uncertainty to the cross section assuming that the
individual uncertainties are uncorrelated.
The procedures and results for the individual quantities considered are
summarized below.
The systematic uncertainties are calculated for the \tauone, \tauthree,
and the combined \tauh analyses separately, with the resulting values
given in Table~\ref{tab:systematic}.

The uncertainty in the cross section due to jet reconstruction is
split into three components: the jet energy scale, its energy
resolution, and its reconstruction efficiency. The uncertainty
from the jet energy scale is calculated by varying the jet energies
according to the uncertainties derived from simulation and the {\em in situ}
calibration using a model containing 22 independent
components~\cite{PERF-2012-01}. The difference between the jet energy
resolution in data and MC simulated events is evaluated by smearing the jet
\pt in the MC sample according to the measured jet resolution in bins of \eta
and \pt~\cite{PERF-2011-04}. The uncertainty in the jet reconstruction
efficiency is evaluated by randomly removing jets according to the
difference in data and MC jet reconstruction efficiencies~\cite{PERF-2012-01}.
The variation in the jet energies is also propagated to the \met calculation.

In the nominal analysis, the $b$-tagging efficiency in simulation is
corrected to agree with data by using \pt- and \eta-dependent
correction factors.
The uncertainty in the correction factors is obtained independently for
$b$-jets, $c$-jets, and light-flavor jets assuming that they are
uncorrelated. The uncertainties of the inefficiency correction factors
that are applied when a jet is not tagged are treated as fully
anticorrelated with the corresponding efficiency correction
factor~\cite{PERF-2012-04}.
This uncertainty is propagated to the cross section by varying the
correction factors by one standard deviation with respect to the central
value.

As in $b$-tagging, correction factors are used to correct for
the difference in
the \tauh-tagging efficiency and the \tauh electron veto efficiency 
between data and simulation. The
uncertainties in the correction factors depend on \pt, \eta, and the \tauh
identification criteria. 
In addition, the \tauh energy scale can affect the final result due
to the \tauh \pt requirement. The energy of the \tauh  is calculated
using MC simulation to correct the observed energy to the true 
energy scale~\cite{PERF-2013-06}. Additional small data-based
corrections are then applied.
The uncertainties due to each of these effects are propagated to
the cross section by varying the correction factors by one standard
deviation.

The systematic uncertainty of \met is evaluated along with the
systematic uncertainty of the associated energy and momentum
of the reconstructed objects as discussed above. Not included in that
calculation are the contributions from low-\pt jets and
energy deposits in the calorimeter cells not associated with a reconstructed
object. This source of uncertainty is evaluated using the difference
between data and simulated $Z\to \mu^+\mu^-$ events containing no jets,
which is similar to the procedure used in Ref.~\cite{PERF-2011-07}.

The systematic uncertainty due to \ttbar modeling is split into two
components. The first is that associated with the choice of ME event generator
and PS/UE event simulation.
The uncertainty associated with the choice of ME event generator is estimated by
comparing the acceptance from the \MCatNLO event generator with that from the
\POWHEG-\textsc{box} event generator. Events from both event generators are processed
through the PS/UE simulator \HERWIG+\JIMMY. The uncertainty due to
the PS on the acceptance is estimated by comparing the \POWHEG+\PYTHIA event generator
to the \POWHEG+\HERWIG event generator.
The second  component of the modeling uncertainty corresponds to the effect
of ISR and FSR on the event selection due to possible extra jets and changes
in the kinematics of the final-state particles and jets. The
nominal \ttbar sample is compared to samples with variations
of the renormalization and factorization scales 
and the regularization parameter as described in Sec.~\ref{sec:samples}.

The systematic uncertainties due to the various backgrounds that contain
real \tauh, are derived using the MC samples described in
Sec.~\ref{sec:samples} and
the uncertainties of the theoretical cross sections. The two
largest sources of real \tauh backgrounds are single-top and $W+\jets$
events. All other background contributions to the systematic uncertainty
are negligible.
For single-top, the uncertainty in the 
cross section of the MC sample is varied by one standard deviation
and propagated to the cross section. For the
$W+\jets$ background, the same procedure is followed but
is validated
using a method based on the $W$-boson charge asymmetry in data
as described in Refs.~\cite{TOPQ-2011-08, TOPQ-2011-07, TOPQ-2012-18},
which gives agreement with the estimation based on the theoretical
uncertainty.

To estimate the systematic uncertainty in the number of misidentified
\tauh, the effect of variations of the main components
of this analysis are examined. The main components are:
\begin{enumerate*}[{label=(\arabic*)}]
  \item the MC-based background subtraction of the real \tauh,
  \item uncertainty in the flavor dependence of the \FF,
  \item uncertainty associated with the $\eta$-$\pt$ binning of
    the \FF.
\end{enumerate*}
In calculating the \FF,
the largest contribution from real \tauh is from $Z+\jets$ events, 
as the final state
$Z\to\tau^+\tau^-\to \tauh\;\mu+X$ satisfies the selection.
To estimate this component of the uncertainty, the
$Z+\jets$ cross
section is varied by $\pm 1$ standard deviation. This variation leads
to an average 
uncertainty of $5\%$  over the \pt-\eta range for this component of
the \FF.
The \FF is calculated in a sample dominated by light-flavor jets.
To estimate the systematic uncertainty of the flavor composition,
the \FF is also derived in a gluon-jet-dominated sample with four jets
and low \met. 
Using this sample the \FF  is calculated and applied to the
signal sample, resulting in an  uncertainty of 20\% in the number of
misidentified \tauh events. 
Since the \FF is calculated in \pt-\eta bins, the bin size is also varied
to estimate the uncertainty in the final result. The uncertainty in the
final result is found
to be approximately $5\%$ of the calculated number of misidentified
\tauh events.

The absolute luminosity scale is derived from beam-separation scans
performed in November 2012.
From the calibration of the absolute luminosity scale,
the uncertainty in the total integrated luminosity is evaluated
following the procedure described in Ref.~\cite{DAPR-2013-01} and
is found to be 1.9\%. This uncertainty is then propagated to the
cross section measurements yielding a 2.3\% uncertainty, which is
reported independent of the other systematic uncertainties. 
\begin{table}[htbp]
  \centering
  \caption{Relative percent uncertainties in the measured cross
    section in the \tauone, \tauthree and combined \tauone and
    \tauthree (\tauh) final states. In the \tauone (\tauthree)
    analysis, all \tauh in the event are required to be \tauone
    (\tauthree). For the combined analysis, the \tauh in an
    event could be of either type.}
  \label{tab:systematic}
  \begin{tabular}{lccc}
    \toprule
    Uncertainty & \tauone &
     \tauthree & \tauh\\
    \midrule
Total Systematic & - 11 /+ 11 & - 16 /+ 14 & - 12 /+ 12 \\
\midrule
Jet energy scale & - 4.0 /+ 4.2 & - 8.4 /+ 5.7 & - 5.0 /+ 4.5 \\
$b$-tag efficiency & - 4.7 /+ 5.0 & - 4.8 /+ 5.0 & - 4.7 /+ 5.0 \\
$c$-mistag efficiency & - 1.6 /+ 1.6 & - 1.5 /+ 1.5 & - 1.6 /+ 1.6 \\
Light-jet mistag efficiency & - 0.3 /+ 0.3 & - 0.5 /+ 0.5 & - 0.4 /+ 0.4 \\
\met & - 0.3 /+ 0.5 & - 1.7 /+ 0.5 & - 0.6 /+ 0.4 \\
\tauh identification & - 3.5 /+ 3.4 & - 6.0 /+ 5.6 & - 4.1 /+ 3.9 \\
\tauh energy scale & - 2.1 /+ 2.0 & - 1.2 /+ 1.4 & - 1.9 /+ 1.9 \\
Jet vertex fraction & - 0.1 /+ 0.3 & - 0.3 /+ 0.3 & - 0.2 /+ 0.3 \\
Jet energy resolution & - 1.4 /+ 1.4 & - 0.2 /+ 0.2 & - 1.1 /+ 1.1 \\
Generator & - 1.5 /+ 1.5 & - 2.5 /+ 2.5 & - 2.1 /+ 2.1 \\
Parton Shower & - 2.0 /+ 2.0 & - 2.6 /+ 2.6 & - 2.1 /+ 2.1 \\
ISR/FSR & - 6.2 /+ 6.2 & - 8.5 /+ 8.5 & - 6.7 /+ 6.7 \\
Misidentified-\tauh background & - 1.3 /+ 1.4 & - 2.0 /+ 2.2 & - 1.6 /+ 1.6 \\
$W+\jets$ background & - 2.9 /+ 2.9 & - 3.6 /+ 3.6 & - 3.0 /+ 3.0 \\
\midrule
Statistics & - 2.2 /+ 2.2 & - 5.6 /+ 5.6 & - 1.7 /+ 1.7 \\
\midrule
Luminosity & - 2.3 /+ 2.3 & - 2.3 /+ 2.3 & - 2.3 /+ 2.3 \\
    \bottomrule
  \end{tabular}
\end{table}
\section{Results and Interpretation}
\label{sec:result}
\begin{figure}[htb]
\subfloat[]{
  \includegraphics[width=0.47\textwidth]{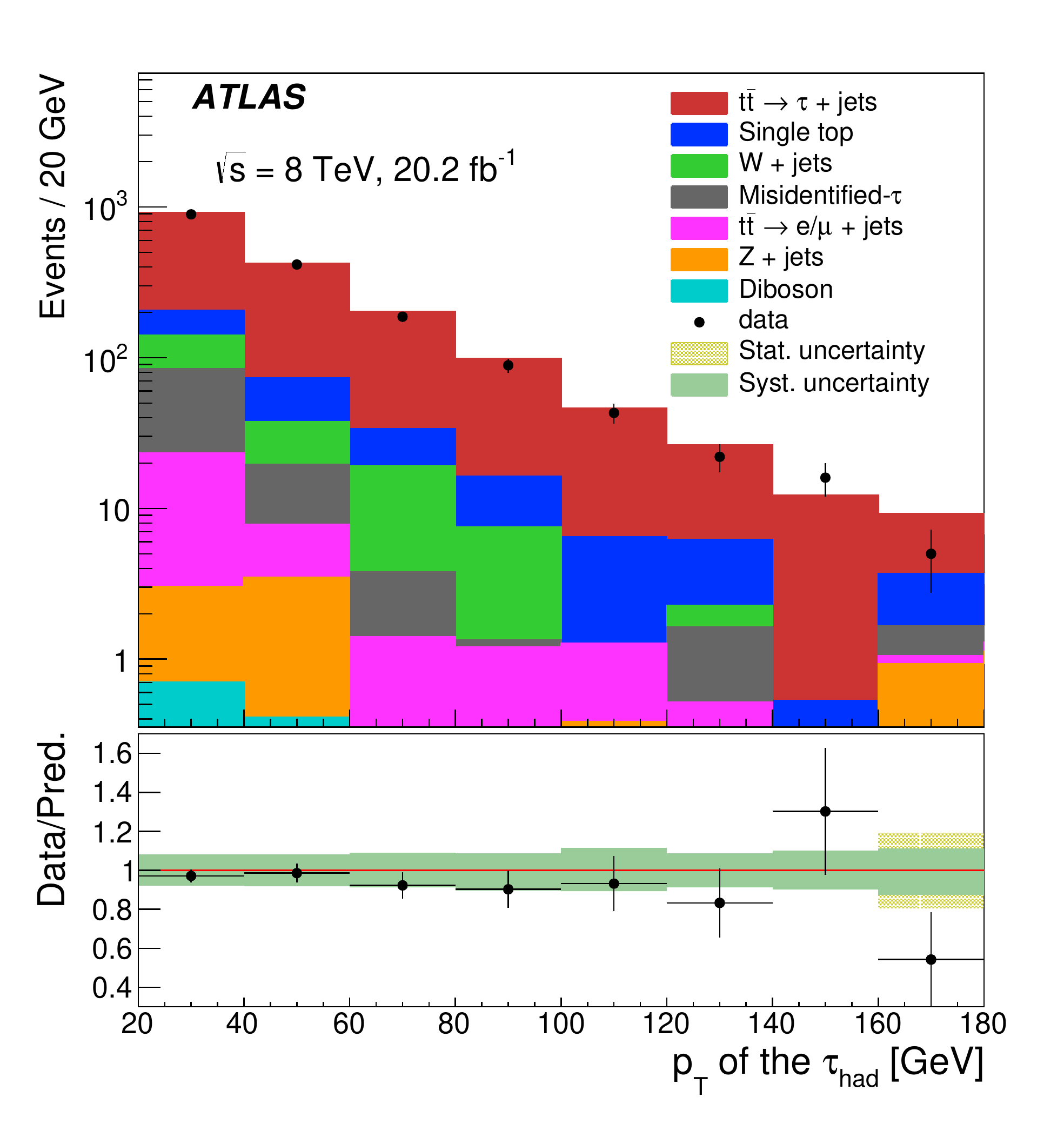}
\label{fig:taupt0}
}
\subfloat[]{
  \includegraphics[width=0.47\textwidth]{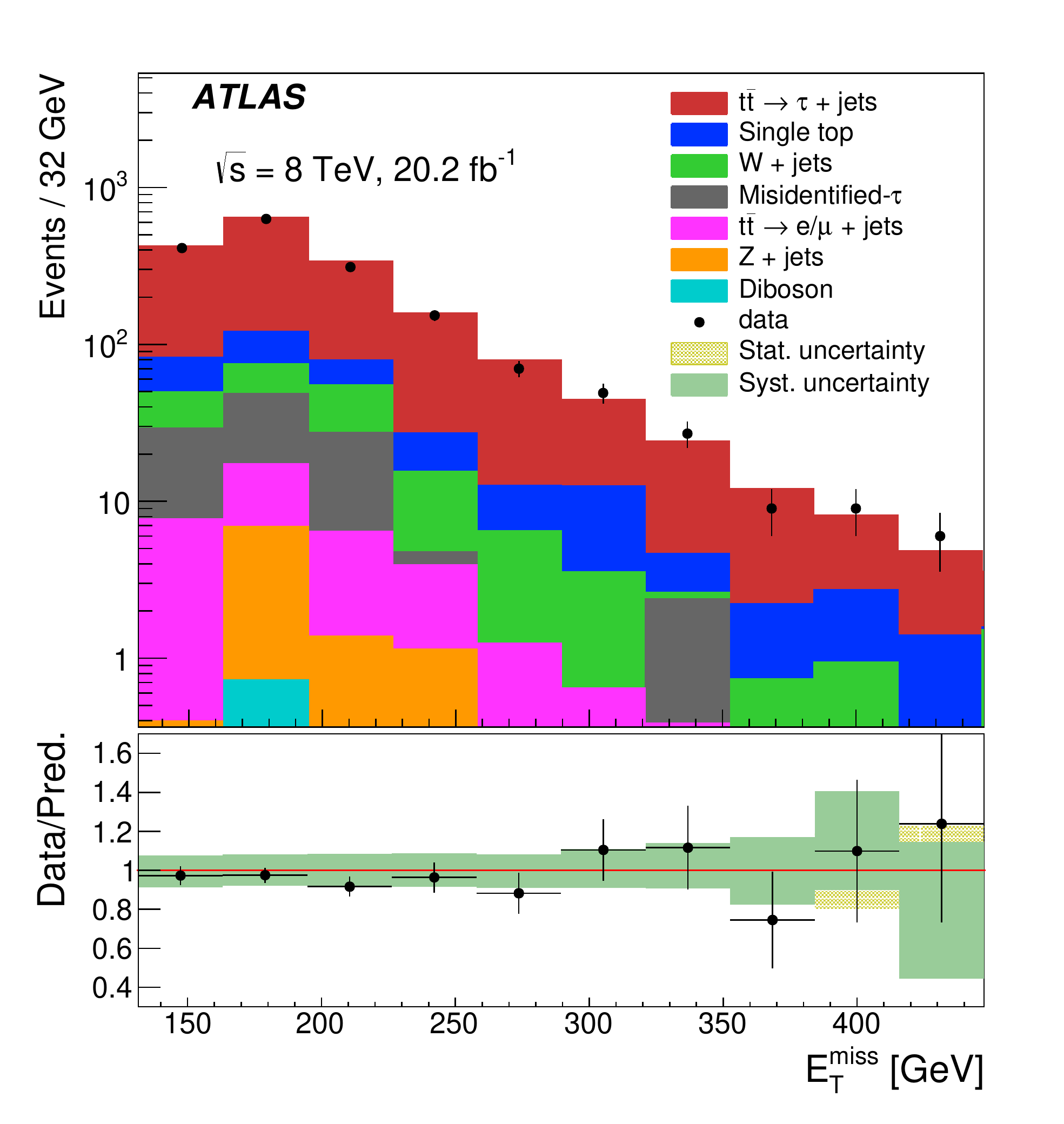}
\label{fig:met}
}
  \caption{ The distribution of the \protect \subref{fig:taupt0} \pt of the \tauh having
    highest transverse momentum in the event 
    and \protect \subref{fig:met} the missing transverse momentum,
    \met. The observed data are compared to the predictions.}
  \label{fig:result}
\end{figure}

The number of events observed for each \tauh type and for the combined
analysis are reported in Table~\ref{tab:yields} along with the predicted
number of background events. The uncertainties associated with the
cross-section
measurement from each of the different sources are reported in
Table~\ref{tab:systematic}.
Figure~\ref{fig:result}
shows the kinematic distributions of the predicted background and
signal processes with the observed data  superimposed, where 
the signal-to-background ratio is approximately 4:1. 

The cross sections for each \tauh type measured separately are
\begin{equation*}
    \begin{aligned}
    &\sigma_{\ttbar}(\tauone+\jets)= 237\pm 5 {\text{(stat)}}\pm 26 {\text{(syst)}}\pm 5{\text{(lumi)}}{\ \si{\pb}},\\
    &\sigma_{\ttbar}(\tauthree+\jets)= 243\pm 14 {\text{(stat)}}^{+34}_{-38} {\text{(syst)}}\pm 6{\text{(lumi)}}{\ \si{\pb}},
  \end{aligned}
\end{equation*}
and the cross section for the combined analysis is
\begin{equation*}
  \sigma_{\ttbar}= 239 \pm  {4\text{(stat)}}\pm 28 {\text{(syst)}}
  \pm 5{\text{(lumi)}}{\ \si{pb}}.
\end{equation*}
The combined cross section has an uncertainty of 12\% and is in agreement
with the previous
measurements of the ATLAS Collaboration for the $e+\jets$ and $\mu+\jets$
final states~\cite{TOPQ-2013-06}. Since the analysis is
performed at a fixed top-quark mass, samples are generated at various
masses to study the dependence of the measured cross section on \mtop.
The variation is found to be
$(\Delta\sigma/\sigma)/\Delta\mtop=-2.6\%\si{\gev^{-1}}$.

In order to quantify the compatibility of this result with the SM
and explore the allowed range for non-SM processes,
a frequentist significance test using a background-only
hypothesis is used to compare the observed number
of events with the SM prediction. 
In this procedure, the $\ttbar\to\tau+X$ process is considered a
background and estimated according to the SM prediction taking into account
the corresponding uncertainty. This statistical analysis is also used to
derive a  limit in a model-independent
manner on possible beyond-the-SM (BSM) physics.
A confidence level for the background-only
hypothesis ($\text{CL}_b$) of 0.48 corresponding to a $p$-value of
0.52 is observed, which indicates good agreement between the observed data
and the SM processes. An upper limit at 95\% confidence level (CL) on
the number of BSM events is derived using the $\text{CL}_s$ likelihood
ratio method described in Ref.~\cite{Read:2002hq}. The upper
limit is calculated with the observed number of events, the expected
background, and the background uncertainty.  Dividing the upper limits
on the number of BSM events by the integrated luminosity of the data
sample, the resulting value can be interpreted as the upper limit on the
visible BSM cross section, $\sigma_{\text{vis}} = \sigma \times
\epsilon$, where $\sigma$ ($\epsilon$) is the production cross
section (efficiency) for the BSM process. Table~\ref{tab:MIUL} summarizes
the observed
number of events, the estimated SM background yield, and the expected and
observed upper limits on the event yields and on the
$\sigma_{\text{vis}}$ from any BSM process. The efficiency for each
SM process used to calculate this limit is reported in Table~\ref{tab:Aeff}.

\begin{table}[htbp]
\begin{center}
\caption{\label{tab:MIUL}
Limits on possible BSM events in this sample.
Top to bottom: Number of observed events, expected SM processes yield, 
95\% CL
observed (expected) upper limits on the number of BSM events and the visible cross section
($\langle\epsilon\sigma\rangle_{\text{obs. (exp.)} }^{95}$).
}
\bgroup
\def\arraystretch{2}%
\begin{tabular}{lr}
\toprule
Observed data & 1678 \\
Expected SM background  &  1751 $\pm$ 42 \\
$S_{\text{obs. (exp.)} }^{95}$ &  $446$ $\left({444}^{+40}_{-21}\right)$ \\
$\langle\epsilon \sigma\rangle_{\text{obs.(exp.)}}^{95}$[fb] & $22$ $\left({22}^{+2}_{-1}\right)$ \\
\bottomrule
\end{tabular}
\egroup
\end{center}
\end{table}

\begin{table}[htbp]
\begin{center}
\caption{\label{tab:Aeff}
The efficiency for each SM process estimated in
simulation.}
\bgroup
\def\arraystretch{1.4}%
\begin{tabular}{c r}
\toprule
Process &  Efficiency ($\epsilon$) \\
\toprule
$\ttbar \rightarrow \tau+\jets$ & $5.0\times10^{-4}$ \\
$\ttbar \rightarrow e/\mu+\jets $ &  $1.0\times10^{-5}$\\  
Single top & $1.6\times10^{-4}$ \\
W+jets & $3.7\times10^{-7}$ \\
Z+jets & $2.4\times10^{-7}$ \\
Diboson & $2.8\times10^{-6}$ \\
\bottomrule
\end{tabular}
\egroup
\end{center}
\end{table}

Using the same data sample as the cross-section measurement,
an upper limit on the flavor changing
process $t\to qH\to q\tau^+\tau^-$ is set by performing a modified analysis
and then calculating a limit in a manner that is similar to that of
the model-independent limit. In the modified analysis, 
exactly one identified $b$-jet and two \tauh are required. Performing the
same statistical analysis as for the cross-section measurement, a
95\%  CL observed (expected)  upper limit of 0.6\% (0.9\%) is set
on the $\BR(t\to qH) \times \BR(H\to\tau \tau)$. At present, this is
the only analysis that can explore the channel $t\to qH\to q\tau\tau$
and, hence, is the first search using the $H\to\tau\tau$ final state. 
Assuming the SM $\BR(H \to \tau \tau)=6\%$, the 95\% CL
observed (expected) upper limit set on the $\BR(t\to qH)$ is 10\% (15\%).  
A dedicated ATLAS measurement achieves a 95\% CL
upper limit of 0.45\% on the $\BR(t\to qH)$ in the combination of
Higgs boson final states
$H \rightarrow bb$, $H \to \gamma \gamma$ and $H \to
\text{multilepton}\ (e,\mu) $~\cite{TOPQ-2014-14}. 
\section{Summary}
\label{sec:conclusion}
A measurement of the $pp\to \ttbar+X$ cross section at
$\sqrt{s}=\SI{8}{\tev}$ using \SI{20.2}{\per\fb} of integrated luminosity
collected with the ATLAS detector has been performed in the
$\ttbar\to\tau\nu_{\tau} q\bar{q}' b\bar{b}$ final state using hadronic
decays of the \tau lepton.
The cross section is  measured
separately for hadronic decays of the \tau lepton into one or three
charged particles. 
A single analysis using a combination of both decay modes is also performed.
The cross section measured in the single analysis is
$\sigma_{\ttbar}= 239 \pm  {4\text{(stat)}}\pm 28 {\text{(syst)}}
\pm 5{\text{(lumi)}}{~\si{pb}}$, assuming a
top-quark mass of $\mtop = 172.5~\text{GeV}$. The measured cross section is
in agreement with the SM prediction of  $253^{+13}_{-15}\
~\si{\pb}$. A
statistical analysis is performed to check the consistency of the
observed number of events in data with the predicted number of events
from various SM processes. Following a frequentist approach,
the confidence level observed with the SM-only hypothesis is 0.48
and the calculated $p$-value is 0.52, which indicates good agreement of
the SM prediction with the observed data. A model-independent upper limit
on the visible cross section for any non-SM process
is also calculated. The observed (expected)
upper limit at 95\% confidence level on the visible cross section of any
non-SM processes is $22\left(22^{+2}_{-1}\right)$
$\text{fb}$.
\section*{Acknowledgements}
We thank CERN for the very successful operation of the LHC, as well as the
support staff from our institutions without whom ATLAS could not be
operated efficiently.

We acknowledge the support of ANPCyT, Argentina; YerPhI, Armenia; ARC, Australia; BMWFW and FWF, Austria; ANAS, Azerbaijan; SSTC, Belarus; CNPq and FAPESP, Brazil; NSERC, NRC and CFI, Canada; CERN; CONICYT, Chile; CAS, MOST and NSFC, China; COLCIENCIAS, Colombia; MSMT CR, MPO CR and VSC CR, Czech Republic; DNRF and DNSRC, Denmark; IN2P3-CNRS, CEA-DSM/IRFU, France; GNSF, Georgia; BMBF, HGF, and MPG, Germany; GSRT, Greece; RGC, Hong Kong SAR, China; ISF, I-CORE and Benoziyo Center, Israel; INFN, Italy; MEXT and JSPS, Japan; CNRST, Morocco; FOM and NWO, Netherlands; RCN, Norway; MNiSW and NCN, Poland; FCT, Portugal; MNE/IFA, Romania; MES of Russia and NRC KI, Russian Federation; JINR; MESTD, Serbia; MSSR, Slovakia; ARRS and MIZ\v{S}, Slovenia; DST/NRF, South Africa; MINECO, Spain; SRC and Wallenberg Foundation, Sweden; SERI, SNSF and Cantons of Bern and Geneva, Switzerland; MOST, Taiwan; TAEK, Turkey; STFC, United Kingdom; DOE and NSF, United States of America. In addition, individual groups and members have received support from BCKDF, the Canada Council, CANARIE, CRC, Compute Canada, FQRNT, and the Ontario Innovation Trust, Canada; EPLANET, ERC, ERDF, FP7, Horizon 2020 and Marie Sk{\l}odowska-Curie Actions, European Union; Investissements d'Avenir Labex and Idex, ANR, R{\'e}gion Auvergne and Fondation Partager le Savoir, France; DFG and AvH Foundation, Germany; Herakleitos, Thales and Aristeia programmes co-financed by EU-ESF and the Greek NSRF; BSF, GIF and Minerva, Israel; BRF, Norway; CERCA Programme Generalitat de Catalunya, Generalitat Valenciana, Spain; the Royal Society and Leverhulme Trust, United Kingdom.

The crucial computing support from all WLCG partners is acknowledged gratefully, in particular from CERN, the ATLAS Tier-1 facilities at TRIUMF (Canada), NDGF (Denmark, Norway, Sweden), CC-IN2P3 (France), KIT/GridKA (Germany), INFN-CNAF (Italy), NL-T1 (Netherlands), PIC (Spain), ASGC (Taiwan), RAL (UK) and BNL (USA), the Tier-2 facilities worldwide and large non-WLCG resource providers. Major contributors of computing resources are listed in Ref.~\cite{ATL-GEN-PUB-2016-002}.
\printbibliography
\newpage
\begin{flushleft}
{\Large The ATLAS Collaboration}

\bigskip

M.~Aaboud$^\textrm{\scriptsize 137d}$,
G.~Aad$^\textrm{\scriptsize 88}$,
B.~Abbott$^\textrm{\scriptsize 115}$,
J.~Abdallah$^\textrm{\scriptsize 8}$,
O.~Abdinov$^\textrm{\scriptsize 12}$,
B.~Abeloos$^\textrm{\scriptsize 119}$,
O.S.~AbouZeid$^\textrm{\scriptsize 139}$,
N.L.~Abraham$^\textrm{\scriptsize 151}$,
H.~Abramowicz$^\textrm{\scriptsize 155}$,
H.~Abreu$^\textrm{\scriptsize 154}$
% ...
\bigskip
\\
% ...
$^{8}$ Department of Physics, The University of Texas at Arlington, Arlington TX, United States of America\\
% ...
$^{12}$ Institute of Physics, Azerbaijan Academy of Sciences, Baku, Azerbaijan\\
% ...
$^{88}$ CPPM, Aix-Marseille Universit{\'e} and CNRS/IN2P3, Marseille, France\\
% ...
$^{115}$ Homer L. Dodge Department of Physics and Astronomy, University of Oklahoma, Norman OK, United States of America\\
% ...
$^{119}$ LAL, Univ. Paris-Sud, CNRS/IN2P3, Universit{\'e} Paris-Saclay, Orsay, France\\
% ...
$^{137}$ $^{(a)}$ Facult{\'e} des Sciences Ain Chock, R{\'e}seau Universitaire de Physique des Hautes Energies - Universit{\'e} Hassan II, Casablanca; $^{(b)}$ Centre National de l'Energie des Sciences Techniques Nucleaires, Rabat; $^{(c)}$ Facult{\'e} des Sciences Semlalia, Universit{\'e} Cadi Ayyad, LPHEA-Marrakech; $^{(d)}$ Facult{\'e} des Sciences, Universit{\'e} Mohamed Premier and LPTPM, Oujda; $^{(e)}$ Facult{\'e} des sciences, Universit{\'e} Mohammed V, Rabat, Morocco\\
% ...
$^{139}$ Santa Cruz Institute for Particle Physics, University of California Santa Cruz, Santa Cruz CA, United States of America\\
% ...
$^{151}$ Department of Physics and Astronomy, University of Sussex, Brighton, United Kingdom\\
% ...
$^{154}$ Department of Physics, Technion: Israel Institute of Technology, Haifa, Israel\\
$^{155}$ Raymond and Beverly Sackler School of Physics and Astronomy, Tel Aviv University, Tel Aviv, Israel\\
% ...
$^{a}$ Also at Department of Physics, King's College London, London, United Kingdom\\
$^{b}$ Also at Institute of Physics, Azerbaijan Academy of Sciences, Baku, Azerbaijan\\
$^{c}$ Also at Novosibirsk State University, Novosibirsk, Russia\\
$^{d}$ Also at TRIUMF, Vancouver BC, Canada\\
$^{e}$ Also at Department of Physics {\&} Astronomy, University of Louisville, Louisville, KY, United States of America\\
% ...
\end{flushleft}

\end{document}